\documentstyle{article}
\begin{document}

\baselineskip 6mm
\def\thefootnote{\#}

\newcommand{\mpi}{M^{2}_{\pi^{\pm}}}
\newcommand{\cp}{M^{2}_{\pi}}
\newcommand{\mpin}{M^{2}_{\pi^{\rm o}}}
\newcommand{\mkaon}{M^{2}_{K^{\pm}}}
\newcommand{\ck}{M^{2}_{K}}
\newcommand{\mkanull}{M^{2}_{K^{\rm o}}}
\newcommand{\deltapi}{\Delta M^{2}_{\pi}}
\newcommand{\deltaka}{\Delta M^{2}_{K}}
\newcommand{\circsqm}{\hat{M}^2}
\newcommand{\circpin}{\hat{M}^{2}_{\pi^{\rm o}}}
\newcommand{\circp}{\hat{M}^{2}_{\pi}}
\newcommand{\circpi}{\hat{M}^{2}_{\pi^{\pm}}}
\newcommand{\circkanull}{\hat{M}^{2}_{K^{\rm o}}}
\newcommand{\circk}{\hat{M}^{2}_{K}}
\newcommand{\circkaon}{\hat{M}^{2}_{K^{\pm}}}
\newcommand{\deltacircpi}{\Delta\hat{M}^{2}_{\pi}}
\newcommand{\deltacircka}{\Delta\hat{M}^{2}_{K}}
\newcommand{\mh}{\hat{m}}
\newcommand{\beq}{\begin{equation}}
\newcommand{\beqn}{\begin{eqnarray}}
\newcommand{\eq}{\end{equation}}
\newcommand{\eqn}{\end{eqnarray}}
\newcommand{\no}{\nonumber\\}
\newcommand{\uy}{\bar{U}^{+}}
\newcommand{\qr}{Q_{R}}
\newcommand{\ql}{Q_{L}}
\newcommand{\dou}{d^{\mu}\bar{U}}
\newcommand{\donu}{d^{\nu}\bar{U}}
\newcommand{\duu}{d_{\mu}\bar{U}}
\newcommand{\dunu}{d_{\nu}\bar{U}}
\newcommand{\douy}{d^{\mu}\bar{U}^{+}}
\newcommand{\donuy}{d^{\nu}\bar{U}^{+}}
\newcommand{\duuy}{d_{\mu}\bar{U}^{+}}
\newcommand{\xy}{\chi^{+}}
\newcommand{\xyu}{\chi^{+}\bar{U}}
\newcommand{\uyx}{\bar{U}^{+}\chi}
\newcommand{\xuy}{\chi \bar{U}^{+}}
\newcommand{\uxy}{\bar{U}\chi^{+}}
\newcommand{\frmu}{G^{R}_{\mu}}
\newcommand{\frnu}{G^{R}_{\nu}}
\newcommand{\flmu}{G^{L}_{\mu}}
\newcommand{\flnu}{G^{L}_{\nu}}
\newcommand{\fimu}{G^{I}_{\mu}}
\newcommand{\finu}{G^{I}_{\nu}}
\newcommand{\fro}{G^{R\,\mu\nu}}
\newcommand{\fru}{G^{R}_{\mu\nu}}
\newcommand{\flo}{G^{L\mu\nu}}
\newcommand{\flu}{G^{L}_{\mu\nu}}
\newcommand{\fiu}{G^{I}_{\mu\nu}}
\newcommand{\frgmu}{F^{R}_{\mu}}
\newcommand{\flgmu}{F^{L}_{\mu}}
\newcommand{\figmu}{F^{I}_{\mu}}
\newcommand{\fignu}{F^{I}_{\nu}}
\newcommand{\frgo}{F^{R\mu\nu}}
\newcommand{\frgu}{F^{R}_{\mu\nu}}
\newcommand{\flgo}{F^{L\mu\nu}}
\newcommand{\flgu}{F^{L}_{\mu\nu}}
\newcommand{\figu}{F^{I}_{\mu\nu}}
\newcommand{\coqr}{c^{R\mu}Q_{R}}
\newcommand{\cuqr}{c^{R}_{\mu}Q_{R}}
\newcommand{\coql}{c^{L\mu}Q_{L}}
\newcommand{\cuql}{c^{L}_{\mu}Q_{L}}
\newcommand{\cor}{c^{R\mu}Q}
\newcommand{\cur}{c^{R}_{\mu}Q}
\newcommand{\col}{c^{L\mu}Q}
\newcommand{\cul}{c^{L}_{\mu}Q}
\newcommand{\f}{{\rm F}_{\rm o}^{2}}
\newcommand{\fy}{{\rm F}_{\rm o}^{4}}
\newcommand{\cf}{\frac{{\rm C}}{{\rm F}_{\rm o}^{2}}}
\newcommand{\cfy}{\frac{{\rm C}}{{\rm F}_{\rm o}^{4}}}
\newcommand{\phix}{\varphi}
\newcommand{\laq}{{\cal L}_{2}^{(Q)}}
\newcommand{\laqi}{{\cal L}_{2\;int}^{(Q)}}
\newcommand{\laqr}{{\cal L}_{2}^{(Q)}}
\newcommand{\laqre}{{\cal L}_{E}^{(Q)}}
\newcommand{\laqc}{{\cal L}_{4}^{(Q)}}
\newcommand{\laqrc}{{\cal L}_{4}^{(Q)}}
\newcommand{\mq}{{\cal M}}
\newcommand{\La}{{\cal L}}
\newcommand{\Z}{{\cal Z}}
\newcommand{\D}{{\cal D}}
\newcommand{\bo}{{\rm o}}
\newcommand{\Sa}{{\cal S}}
\newcommand{\A}{\bar{A}}
\newcommand{\bPhi}{\bar{\Phi}}
\newcommand{\F}{\bar{F}}
\newcommand{\U}{\bar{U}}

\pagestyle{empty}

\begin{flushright}
BUTP-94/9
\end{flushright}

\vspace*{1.5cm}

\begin{center}
{\Large {\bf Virtual Photons in Chiral Perturbation Theory}}\footnote{Work
supported in
part by Schweizerischer Nationalfonds.}\\[15mm]
Res Urech\\
Institute for Theoretical Physics, University of Berne\\
Sidlerstr. 5, CH-3012 Berne, Switzerland\\[2mm]
e-mail: urech@itp.unibe.ch\\
[5.5cm]
{\large {\bf Abstract}}\\
\end{center}
In the framework of chiral perturbation theory virtual photons are included.
We calculate the divergences of the generating functional to one loop and
determine the structure of the local action that incorporates the counterterms
which cancel the divergences.
As an application we discuss the corrections to Dashen's theorem at order $e^2
m_q $.

\vspace{4cm}

\hspace{-5mm}May 1994
\newpage

\section{Introduction}

\pagestyle{plain}
\pagenumbering{arabic}

In the chiral limit where the light up, down and strange quark masses go to
zero, the QCD lagrangian  has a $SU(3)_{R}\times SU(3)_{L}$ chiral symmetry
that is
spontaneously broken by the groundstate of the theory .
There are eight Goldstone bosons: The $\pi$'s, $K$'s and $\eta$. Their
interactions are
described by an effective low energy theory, called chiral perturbation
theory (CHPT). At low momenta  the chiral lagrangian can
be expanded in derivatives of the Goldstone fields
and in the masses of the three light quarks. \newline
CHPT is a nonrenormalizable theory: Loops produce
ultraviolet divergences, which can be absorbed by introducing counterterms.
The finite parts of the coupling constants of this counterterm lagrangian are
not
determined by the lagrangian that generated the loops. The complete
information about them is hidden in the QCD lagrangian, but there is
no known way to extract this information from first principle alone. The
couplings  have to be evaluated from experiments.
As shown by Weinberg \cite{weinberg}, the loops are suppressed:
Every loop and the associated counterterm correspond to
successively higher powers of momenta or quark masses. At low energies, the
contributions from higher loops are small.\newline
CHPT is an effective theory: It contains all terms allowed by the symmetry of
the QCD lagrangian in the chiral limit. The local action is generated by the
effective lagrangian, which is in general not determined by the
counterterms alone at higher orders in
the expansion of momenta and quark masses. Additional terms with finite
couplings have to be included. \newline
The generating functional to one loop and the corresponding local action in
the strong sector has been determined by
Gasser and Leutwyler \cite{ga85}. An extension to the $\Delta S=1$ nonleptonic
weak interactions to
one loop has been given by Kambor, Missimer and Wyler \cite{kambor}. Here we
consider
virtual photons and include them in the mesonic sector. We calculate the
divergences of the one-loop functional in presence
of external currents and determine the structure of the
local action at this order.\vspace{-3mm}\newline
\newline
The article is organized as follows.
In the second chapter virtual photons are included in CHPT with three flavours.
We evaluate the divergences of the generating functional to one loop and
determine the structure of the local action at order $p^4$.
In the third chapter we discuss the corrections  to Dashen's theorem
at order $e^2 m_q$ and compare our estimates with the results in the
literature.
In the fourth chapter we summarize our results and in Appendix A the matrix
relations are given that are used to simplify the effective lagrangian at
order $p^4$. Finally, in Appendix B we list the
renormalized masses of the pseudoscalar mesons at order $e^2 m_q $.

\section{Generating Functional}

We suppose that the reader is familiar with CHPT,
otherwise we refer her or him to comprehensive reviews \cite{pich}.\newline
At leading order, the effective lagrangian of the strong and
electromagnetic interactions which respects the
chiral symmetry $SU(3)_{R}\times SU(3)_{L}$, $P$ and $C$ invariance is in the
mesonic sector \cite{ecker}
\begin{eqnarray}\label{effl}
\laq= &-& \frac{1}{4} F_{\mu\nu} F^{\mu\nu} - \frac{\lambda}{2} \left(
{\partial}_{\mu} A^{\mu} \right)^{2} \no
&+& \frac {1}{4} \f < d^{\mu}U^{+} d_{\mu}U + \chi U^{+}+ \chi^{+} U >
+ \mbox{ C} < Q U Q U^{+} >,
\end{eqnarray}
where $F_{\mu\nu}$ is the field strength tensor of the photon field $A_{\mu}$,
$F_{\mu\nu}=\partial _{\mu}A_{\nu}-\partial_{\nu}A_{\mu}$.
$\lambda$ is the gauge fixing parameter and will from now on be kept at
$\lambda =1$. $U$ is an unitary 3$\times 3$-matrix and incorporates the fields
of the eight pseudoscalar mesons,
\beqn \label{para}
UU^{+} ={\bf 1},\hspace{1.9cm}\det U=1,\hspace{6mm}\no
U=exp\,(i\Phi/{\rm F}_{\bo}),\hspace{8mm}
\Phi=\sum_{a=1}^{8}\lambda_{a}\phix_{a}\;,
\eqn
where the $\lambda_{a}$'s are the Gell-Mann matrices, and
\beq \label{phys}
\Phi=\sqrt{2}\left(
\begin{array}{ccc}
\frac{\pi^{\bo}}{\sqrt{2}}+\frac{\eta_{8}}{\sqrt{6}} & \pi^{+} & K^{+}\\[2mm]
\pi^{-} & -\frac{\pi^{\bo}}{\sqrt{2}}+\frac{\eta_{8}}{\sqrt{6}} &
K^{\bo}\\[2mm]
K^{-} & \bar{K}^{\bo} & -\frac{2\eta_{8}}{\sqrt{6}}
\end{array}
\right).
\eq
$d_{\mu}U$ is a covariant derivative, incorporating the couplings to the photon
field $A_{\mu}$,
the external vector and axial vector currents $v_{\mu}$ and $a_{\mu}$,
respectively,
\beq
d_{\mu}U = {\partial}_{\mu} U - i(v_{\mu} + Q A_{\mu} + a_{\mu}) U + i U
(v_{\mu} + Q A_{\mu} - a_{\mu}),
\eq
where $Q$ is the charge matrix of the three light quarks,
\beq
Q=\frac{e}{3}\left(
\begin{array}{ccc}
2&&\\
&-1&\\
&&-1\\
\end{array}\right)
=\frac{e}{2}\left(\lambda_{3}+\frac{1}{\sqrt{3}}\lambda_{8}\right).
\eq
We do not consider singlet vector and axial vector currents and we put
therefore $\mbox{tr }v_{\mu}=\mbox{tr }a_{\mu}=0$. $\chi$ denotes the coupling
of the mesons to the scalar and pseudoscalar currents $s$ and $p$,
respectively,
\beq
\chi=2B_{\bo}(s+ip),
\eq
where $s$ incorporates the mass matrix of the quarks,
\beq
s=\mq+\cdots=\left(
\begin{array}{ccc}
m_{u}&&\\
&m_{d}&\\
&&m_{s}
\end{array}\right)+\cdots
\eq
${\rm F}_{\bo}$ is the pion
decay constant in the chiral limit, ${\rm F}_{\pi}={\rm F}_{\bo} \left[1+
O(m_{q})\right]$.
$B_{\bo}$ is related to the quark condensate $<\bo|\bar{u}u|\bo>= -\f B_{\bo}
\left[1+ O(m_{q})\right]$, and C determines the purely electromagnetic part of
 the masses of the charged pions and kaons in the chiral limit,
\beq
\mpi=\mkaon=2e^{2}\cf + O(m_{q}).
\eq
To ensure the chiral $SU(3)_{R}\times SU(3)_{L}$ symmetry of the lagrangian
$\laq$
we introduce local spurions $\qr,\ql$ instead of the charge matrix
$Q$. The lagrangian is modified,
\beqn\label{efflagr}
\laqr= &-& \frac{1}{4} F_{\mu\nu} F^{\mu\nu} - \frac{1}{2} \left(
{\partial}_{\mu} A^{\mu} \right)^{2} \no
&+& \frac {1}{4} \f < d^{\mu}U^{+}d_{\mu}U + \chi U^{+} + \chi^{+}U >
+ \mbox{ C} < \qr U \ql U^{+} >,
\eqn
and $d_{\mu}U$ is changed to
\beq
d_{\mu}U = {\partial}_{\mu} U - i(v_{\mu} + \qr A_{\mu} + a_{\mu}) U + i U
(v_{\mu} + \ql A_{\mu} - a_{\mu}).
\eq
The rules of chiral transformation are \cite{ga85,ecker}
\beqn
U&\rightarrow& g_{R}Ug^{+}_{L},\no
Q_{I}&\rightarrow& g_{I}Q_{I}g^{+}_{I},\hspace{2cm}I=R,L\no
v_{\mu} + \qr A_{\mu} + a_{\mu}&\rightarrow& g_{R}(v_{\mu} + \qr A_{\mu} +
a_{\mu}) g^{+}_{R} + ig_{R}\partial_{\mu}g^{+}_{R},\no
v_{\mu} + \ql A_{\mu} - a_{\mu}&\rightarrow& g_{L}(v_{\mu} + \ql A_{\mu} -
a_{\mu})g^{+}_{L} +ig_{L}\partial_{\mu}g^{+}_{L},\no
s+ip&\rightarrow& g_{R}(s+ip)g^{+}_{L},\no
g_{R,L}&\in& SU(3)_{R,L}\;.
\eqn
Since the currents $v_{\mu}$ and $a_{\mu}$ count as $O(p)$, where $p$
means the momentum of the external fields, the term $QA_{\mu}$ has the same
dimension. We put
\beq\label{conv}
dim(Q_{R,L})=O(p),\hspace{2cm}dim(A_{\mu})=O(1),
\eq
the lagrangian $\laqr$ is therefore of order $p^{2}$. The convention
(\ref{conv}) has the advantage that electromagnetic interactions do not turn
upside
down the usual chiral counting.
The procedure to obtain the generating functional at order $O(p^4)$ is very
similar to conventional CHPT. The one-loop graphs generated by $\laq$ are of
order $O(p^4)$. They contain divergences which are absorbed by adding tree
graphs, evaluated with the lagrangian $\laqc$ of order $O(p^4)$, see below.
The generating functional becomes, up to and including terms of order $O(p^4)$,
\beq
e^{i\Z(v_{\mu},a_{\mu},s,p)}=N\int\left[dU\right]\left[dA_{\mu}\right]
e^{i\int d^{4}x \left\{\laq+\laqc\right\}},
\eq
where the integration over the fields is carried out in the one-loop
approximation. [Here and below we disregard contributions from the anomaly
altogether.] To evaluate the divergent part of the one-loop
functional, we transform the lagrangian $\laqr$
(\ref{efflagr}) to Euclidean spacetime,
$\laqr\rightarrow\laqre$,  and expand
the fields $U$ and $A_{\mu}$ around the classical
solutions $(\bar{U},\bar{A}_{\mu})$ of the equations of motion,
\beqn\label{fluq}
U&=&ue^{i\xi/{\rm F}_{\bo}}u=u\left({\bf 1}+i\frac{\xi}{{\rm F}_{\bo}}
-\frac{1}{2}\frac{\xi^{2}}{\f}+\cdots\right)u\no
&=&\bar{U}+\frac{i}{{\rm F}_{\bo}}u\xi u-\frac{1}{2\f}u\xi^{2}u+\cdots\no[2mm]
A_{\mu}&=&\bar{A}_{\mu}+\epsilon_{\mu},
\eqn
where we put $\bar{U}=u^{2}$, and $\xi$ is a traceless hermitean matrix.
 We insert the expansion (\ref{fluq}) in the action $\Sa_E$,
\beqn
\Sa_E&=&\int d^{4}x_{E}\;\bar{\La}_{E}^{(Q)}\no[1mm]
&&+\int d^{4}x_{E}\left\{\frac{1}{4}<D_{\mu}\xi D_{\mu}\xi
-[\Delta_{\mu},\xi][\Delta_{\mu},\xi]+\sigma\xi^{2}>\right.\no[1mm]
&&\hspace{1.8cm}-\left.\frac{1}{8}\cf<[H_{R}+H_{L},\xi][H_{R}-H_{L},\xi]>
\right.\no[1mm]
&&\hspace{1.8cm}+\left.\frac{1}{2}{\rm F_{0}}<\left(\xi
[H_{R},\Delta_{\mu}]-H_L D_{\mu}\xi \right)>\epsilon_{\mu}\right.\no[1mm]
&&\hspace{1.8cm}+\left.\frac{1}{2}\epsilon_{\mu}\left(-\partial_{\nu}
\partial_{\nu}+\frac{1}{2}\f <H_{L}^{2}>\right)\epsilon_{\mu}\right\}+\cdots,
\eqn
where the ellipsis denotes higher order terms in the fluctuations $\xi$ and
$\epsilon_{\mu}$. We used the expressions (see also ref.\cite{ga85})
\beqn\label{defi}
D_{\mu}\xi&=&\partial_{\mu}\xi +[\Gamma_{\mu},\xi],\no[2mm]
\Gamma_{\mu}&=&\frac{1}{2}[u^{+},\partial_{\mu}u]-\frac{1}{2}iu^{+}\frmu u-
\frac{1}{2}iu\flmu u^{+},\no[1mm]
\Delta_{\mu}&=&\frac{1}{2}u^{+}d_{\mu}\bar{U}u^{+}=-\frac{1}{2}ud_{\mu}
\bar{U}^{+}u,\no[2mm]
\frmu&=&v_{\mu} +\qr \A_{\mu} +a_{\mu},\no[2mm]
\flmu&=&v_{\mu} +\ql \A_{\mu} -a_{\mu},\no[2mm]
H_{R}&=&u^{+}\{Q,\bar{U}\}u^{+}=u^{+}\qr u+ u\ql u^{+},\no[2mm]
H_{L}&=&u^{+}[Q,\bar{U}]u^{+}=u^{+}\qr u- u\ql u^{+},\no[2mm]
\sigma&=&\frac{1}{2}(u^{+}\chi u^{+}+u\chi^{+} u).
\eqn
$\Gamma_{\mu}$ and $\Delta_{\mu}$ are antihermitean matrices, whereas
$H_{R,L}$ and $\sigma$ are hermitean ones and $\int
d^{4}x_{E}\,\bar{\La}_{E}^{(Q)}$  represents the classical
action of $\laqre$.
The normal brackets $[\cdots]$ denote the commutator, the curly brackets
$\{\cdots\}$ stand for the anti-commutator. We use the
parametrization $\xi=\sum_{a}\xi^{a}\lambda^{a}$, where the $\lambda^{a}$'s are
the Gell-Mann matrices. We define a new covariant
derivative $\Sigma_{\mu}=D_{\mu}+X_{\mu}$ with
\beqn
\Sigma_{\mu}\xi^{a}&=&D_{\mu}\xi^{a}+X_{\mu}^{a\rho}\epsilon^{\rho}\;,\no
\Sigma_{\mu}\epsilon^{\rho}&=&\partial_{\mu}\epsilon^{\rho}+X_{\mu}^{\rho a}
\xi^{a}\;,
\eqn
where $D_{\mu}\xi^{a}$ is understood as
\beq
D_{\mu}\xi^{a}=\partial_{\mu}\xi^{a}+\Gamma_{\mu}^{ab}\xi^{b}
\eq
and
\beqn
\Gamma_{\mu}^{ab}&=&-\frac{1}{2}<[\lambda^{a},\lambda^{b}]\Gamma_{\mu}>,\no
X_{\mu}^{a\rho}&=&-X_{\mu}^{\rho a}=-\frac{1}{4}{\rm F}_{0}<H_L \lambda^{a}>
\delta^{\rho}_{\mu}.
\eqn
The action becomes at the one-loop level
\beqn
\Sa_E|_{one\;loop}&=&\frac{1}{2}\int d^{4}x_{E}\left\{\xi^{a}\left(
-\Sigma_{\mu}\Sigma_{\mu}\delta^{ab} +\sigma^{ab}\right)\xi^{b}+\xi^{a}
\gamma^{a}_{\mu}\epsilon_{\mu}\right.\no
&&\hspace{1.8cm}\left.+\epsilon_{\mu}\left(-\Sigma_{\nu}\Sigma_{\nu}+
\rho\right) \epsilon_{\mu}\right\},
\eqn
where now
\beqn
\sigma^{ab}&=&-\frac{1}{2}<[\Delta_{\mu},\lambda^{a}][\Delta_{\mu},
\lambda^{b}]>
+\frac{1}{4}<\sigma\{\lambda^{a},\lambda^{b}\}>\no[1mm]
&&-\frac{1}{4}\cf<[H_{R}+H_{L},\lambda^{a}][H_{R}-H_{L},\lambda^{b}]>
-\frac{1}{4}\f<H_L \lambda^{a}><H_L \lambda^{b}>, \no[2mm]
\gamma^{a}_{\mu}&=&{\rm F}_{\bo}<\left([H_{R},\Delta_{\mu}]+
\frac{1}{2}D_{\mu}H_{L}\right)\lambda^{a}>,\no[2mm]
\rho&=&\frac{3}{8}\f<H_{L}^{2}>.
\eqn
We collect the fluctuations of the mesons and of the photon field and define a
new flavour space $\eta^{A}$, where $A$ runs from 1 to 12,
$\eta=(\xi^{1},\ldots ,\xi^{8},\epsilon^{0},\ldots,\epsilon^{3})$. The action
above can now be written as a quadratic form,
\beq
\Sa_E|_{one\;loop}=\frac{1}{2}\int d^{4}x_{E}\; \eta^{A}\left(
-\Sigma_{\mu}\Sigma_{\mu}\delta^{AB}+\Lambda^{AB}\right)\eta^{B},
\eq
where $\Sigma_{\mu}$ and $\Lambda^{AB}$ are $12\times12$-matrices,
\beqn
\Sigma_{\mu}&=&\partial_{\mu}{\bf 1}+
\left( \begin{array}{cc}
\Gamma^{ab}_{\mu}&X_{\mu}^{a\rho}\\
&\\
X_{\mu}^{\sigma b}&0
\end{array} \right)
=\partial_{\mu}{\bf 1}+Y_{\mu},\nonumber\\
\no[2mm]
\Lambda&=&
\left( \begin{array}{cc}
\sigma^{ab}&\frac{1}{2}\gamma^{a\rho}\\
&\\
\frac{1}{2}\gamma^{\sigma b}&\rho\delta^{\sigma\rho}
\end{array} \right) .\\
\nonumber
\eqn
This leads to a gaussian integral for the
generating functional,
\beqn
e^{-\Z_{E}|_{one\;loop}}&=&N\int d\mu[\eta]
e^{-\frac{1}{2}(\eta,\D\eta)}\no
&=&N^{\prime}(\det \D)^{-\frac{1}{2}},
\end{eqnarray}
where we defined the differential operator
$\D^{AB}=-\Sigma_{\mu}\Sigma_{\mu}\delta^{AB}+
\Lambda^{AB}$  and
\begin{equation}
(f,g)=\sum_{A}\int d^{4}x_{E}f^{A}g^{A}.
\end{equation}
Omitting the constant contribution we arrive at
\begin{equation}
\Z_{E}|_{one\;loop}=\frac{1}{2}\ln (\det \D).
\end{equation}
To renormalize the determinant we use dimensional regularization.
In Minkowski spacetime, the one-loop functional in $d=4$ dimensions is
given by \cite{ga84}
\beq\label{zone}
\Z_{one\;loop}=-\frac{1}{16\pi^{2}}\frac{1}{d-4}\int d^{4}x\; \mbox{Tr}\left(
\frac{1}{12}Y_{\mu\nu}Y^{\mu\nu}+\frac{1}{2}\Lambda^{2}\right) + \mbox{ finite
parts},
\eq
where Tr means the trace in the flavour space $\eta^{A}$ and $Y_{\mu\nu}$
denotes the field strength tensor of $Y_{\mu}$,
\beq
Y_{\mu\nu}=\partial_{\mu}Y_{\nu}-\partial_{\nu}Y_{\mu} +[Y_{\mu},Y_{\nu}].
\eq
\begin{table}
\begin{center}
\begin{tabular}{c|cc}
\hspace{3cm}&\hspace{4cm}&\hspace{4cm}\\[-2mm]
& $P$ & $C$ \\[3mm] \hline
&&\\
$U$ & $U^{+}$ & $U^{T}$\\[3mm]
$d_{\mu}U$ & $d^{\mu}U^{+}$ & $(d_{\mu}U)^{T}$\\[3mm]
$\frmu$ & $G^{L\mu}$ & ${G^{L}_{\mu}}^{T}$\\[3mm]
$\flmu$ & $G^{R\mu}$ & ${G^{R}_{\mu}}^{T}$\\[3mm]
$\chi$ & $\chi^{+}$ & $\chi^{T}$\\[3mm]
$\qr$ & $\ql$ & $\ql^{T}$\\[3mm]
$\cuqr$ & $\coql$ & $(\cuql)^{T}$\\[3mm]
$\ql$ & $\qr$ & $\qr^{T}$\\[3mm]
$\cuql$ & $\coqr$ & $(\cuqr)^{T}$
\end{tabular}
\end{center}
\vspace{5mm}
\caption{{\it P and C transformation properties for the fields, the charges and
for their derivatives. The defintion of $\cuqr$ and $\cuql$ is given in the
text. Spacetime arguments are suppressed.}}
\vspace{1cm}
\end{table}
In table 1 we list $P$ and $C$ transformation properties for the fields and
quantities used in the one-loop functional. $\cuqr$ and $\cuql$ mean covariant
derivatives of $\qr$ and $\ql$, respectively,
\beq
c^{I}_{\mu}Q_{I}=\partial_{\mu} Q_{I}-i[G^{I}_{\mu},Q_{I}],\hspace{2cm}I=R,L\;,
\eq
that transform under $SU(3)_{R}\times SU(3)_{L}$ in the same way as $\qr$ and
$\ql$,
\beqn
c^{I}_{\mu}Q_{I}&\rightarrow& g_{I}\;c^{I}_{\mu}Q_{I}\;g_{I}^{+},
\hspace{2cm}I=R,L\no
g_{R,L}&\in& SU(3)_{R,L}\,.
\eqn
The divergences in $\Z_{one\; loop}$ can be absorbed by adding the
effective lagrangian $\laqrc$ that contains all terms of order $O(p^4)$ allowed
by the chiral symmetry, $P$ and $C$ invariance. Since in the
usual physical picture the charge matrix is a constant and has
no chirality, we put again $\qr=\ql=Q$ and $\partial_{\mu}Q=0$.
We keep the notation $c^{R,L}_{\mu}Q=-i[G^{R,L}_{\mu},Q]$ in order to remember
the correct chiral transformation of $c^{I}_{\mu}Q_{I}$. By using the matrix
relations that we list in Appendix A, the lagrangian $\laqc$ can be simplified
to
\beqn\label{counter}
\lefteqn{\laqc =\hspace{2mm}L_{1} < \douy \duu >^{2}\hspace{2mm} +\hspace{2mm}
L_{2}< \douy \donu >< \duuy \dunu >}\no[1mm]
&&+\hspace{2mm}L_{3} < \douy \duu \donuy \dunu >\hspace{2mm}+\hspace{2mm}L_{4}
< \douy \duu >< \xyu + \xuy >\no[1mm]
&&+\hspace{2mm}L_{5} < \douy \duu ( \xyu + \uyx
)>\hspace{2mm}+\hspace{2mm}L_{6} < \xyu + \xuy >^{2}\hspace{2mm}
+\hspace{2mm}L_{7} < \xyu - \xuy >^{2}\no[1mm]
&&+\hspace{2mm}L_{8} < \xyu
\xyu + \xuy \xuy >\hspace{2mm}-\hspace{2mm}iL_{9} < \dou \donuy\fru + \douy
\donu\flu
>\no[1mm]
&&+\hspace{2mm}L_{10} < \fro \U \flu \uy >\hspace{2mm}
+\hspace{2mm}H_{1} < \fro \fru + \flo \flu >\hspace{2mm}+\hspace{2mm}H_{2} <
\xy \chi >\no[1mm]
&&+\hspace{2mm}K_{1} \f < \douy \duu >< Q^{2}>
\hspace{2mm}+\hspace{2mm}K_{2} \f < \douy \duu >< Q \U Q \uy >\no[1mm]
&&+\hspace{2mm}K_{3} \f\left(<\douy Q \U><\duuy Q \U>+<\dou Q\uy><\duu Q\uy>
\right)\no[1mm]
&&+\hspace{2mm}K_{4} \f < \douy Q \U >< \duu Q \uy >+\hspace{2mm}
K_{5} \f < \left( \douy \duu + \dou \duuy \right) Q^{2} >\no[1mm]
&&+\hspace{2mm}K_{6} \f < \douy \duu Q \uy Q \U + \dou \duuy Q \U Q \uy
>\no[1mm]
&&+\hspace{2mm} K_{7} \f< \xyu + \uyx >< Q^{2}
>\hspace{2mm}+\hspace{2mm}K_{8} \f < \xyu + \uyx >< Q \U Q \uy >\no[1mm]
&&+\hspace{2mm} K_{9} \f <(\xyu + \uyx
+ \xuy + \uxy) Q^{2} >\no[1mm]
&&+\hspace{2mm}K_{10} \f <( \xyu +\uyx ) Q \uy Q \U + ( \xuy + \uxy ) Q \U Q
\uy >\no[1mm]
&&+\hspace{2mm} K_{11} \f <( \xyu - \uyx)Q\uy Q\U
+ (\xuy - \uxy ) Q\U Q\uy >\no[1mm]
&&+\hspace{2mm}K_{12} \f < \douy \left[ \cur , Q \right] \U
+ \dou \left[ \cul , Q \right] \uy >\no[1mm]
&&+\hspace{2mm}K_{13} \f < \cor \U \cul \uy >\hspace{2mm}+\hspace{2mm}K_{14} \f
< \cor\; \cur + \col\;\cul > \no[1mm]
&&+\hspace{2mm}K_{15}\fy <Q \U Q\uy>^{2}+\hspace{2mm}K_{16}\fy <Q \U
Q\uy><Q^2>\hspace{2mm}+\hspace{2mm}K_{17}\fy <Q^2>^2.
\eqn
where $\fru$ and $\flu$ are the field strength tensors of $\frmu,\flmu$,
respectively,
\beqn\label{gmu}
\fiu &=& {\partial}_{\mu} \finu - {\partial}_{\nu} \fimu - i \left[ \fimu ,
\finu \right], \hspace{2cm}I=R,L.
\eqn
If we put $Q=0$, the terms with the couplings $K_{i}$ vanish and
we are left with the contribution to $\laqc$ from the purely strong sector,
calculated by Gasser and Leutwyler \cite{ga85}. In (\ref{counter}) we have only
considered contributions made out of the building
blocks in table 1\,. In particular, terms proportional to polynomials in
$\A_{\mu}$ are omitted.
The coupling constants $L_{i},H_{i}$ and $K_{i}$ are defined as
\beqn
L_{i}&=&\Gamma_{i}\lambda +L_{i}^{r}(\mu),\no
H_{i}&=&\Delta_{i}\lambda + H_{i}^{r}(\mu),\no
K_{i}&=&\Sigma_{i} \lambda + K_{i}^{r}(\mu).
\eqn
$\lambda$ contains a pole in $d=4$ dimensions,
\beq
\lambda
=\frac{\mu^{d-4}}{16\pi^{2}}\left\{\frac{1}{d-4}- \frac{1}{2}\left[\ln 4\pi+
\Gamma^{\prime}(1)+1\right]\right\}.
\eq
The coefficients $\Gamma_{i},\Delta_{i}$ and the finite parts
$L_{i}^{r}(\mu)$ are listed
in ref.\cite{ga85}, whereas $H_{1}^{r}$ and $H_{2}^{r}$ are of no physical
significance. The coefficients $\Sigma_{i}$ are
\beq
\begin{array}{llll}
&&\\
\Sigma_{1}=\frac{3}{4}&\Sigma_{2}=Z & \Sigma_{3}=-\frac{3}{4}\\[3mm]
\Sigma_{4}=2Z & \Sigma_{5}=-\frac{9}{4} &
\Sigma_{6}=\frac{3}{2}Z\\[3mm]
\Sigma_{7}=0 & \Sigma_{8}=Z & \Sigma_{9}=-\frac{1}{4}\\[3mm]
\Sigma_{10}=\frac{1}{4}+\frac{3}{2}Z &
\Sigma_{11}=\frac{1}{8} & \Sigma_{12}=\frac{1}{4}\\[3mm]
\Sigma_{13}=0 & \Sigma_{14}=0 & \Sigma_{15}=\frac{3}{2}+3Z
+20Z^{2}\\[3mm]
\Sigma_{16}=-3-\frac{3}{2}Z-4Z^2\hspace{7mm}
& \Sigma_{17}=\frac{3}{2}-\frac{3}{2}Z+2Z^2\hspace{7mm} &\\
&&
\end{array}
\eq
with
\beq
Z=\cfy.
\eq
We choose the renormalized coefficients $K_{i}^{r}(\mu)$ in such a way that
the $K_{i}$ themselves do not depend on the scale $\mu$,
\beq\label{muk}
\mu\frac{d}{d\mu}K_{i}=\Sigma_{i}\frac{\mu^{d-4}}{16\pi^{2}}+\mu
\frac{d}{d\mu}K_{i}^{r}(\mu)+O(d-4)=0.
\eq

\section{Corrections to Dashen's Theorem}

This chapter is separated in four sections. In the first section we calculate
the formal expression for the corrections to
Dashen's theorem at order $e^2 m_q$, i.e. the squared mass difference
$(\mkaon-\mkanull)-(\mpi-\mpin)$. In the second section a numerical estimate
is given with an upper limit on the unknown terms. In the third section  we
give an
independent approach on the basis of ref.\cite{leuti}.
Finally, we compare in the fourth section the obtained values with the results
in the literature.

\subsection{Masses at Order $e^2 m_q$}

In the chiral limit, the masses of the pions and the
kaons are of purely electromagnetic nature. A relation between the squared
masses at order $e^2$ is given by Dashen's theorem \cite{dashen},
\beq\label{dash}
\left(\mpi-\mpin\right)_{e.m.}=\left(\mkaon-\mkanull\right)_{e.m.}
\eq
with
\beqn
\mpi&=&\mkaon=2e^{2}\cf,\no
\mpin&=&\mkanull=0\hspace{2cm}\mbox{ for}\hspace{5mm}m_{q}=0.
\eqn
At lowest order in the quark mass expansion the squared masses of the
pseudoscalar mesons are denoted by $\circsqm_{P}$. We neglect the mass
difference $m_{d}-m_{u}$ and replace $m_{u},m_{d}$ by $\mh=(m_{u}+m_{d})/2$.
If we expand the lagrangian $\laq$ in the parametrization
(\ref{para}) and use the representation (\ref{phys}) of the physical fields,
the squared masses are
\beqn\label{mass}
\circpi&=&2e^{2}\cf+2\hat{m}B_{\bo}\,,\no
\circpin\hspace{1mm}&=&\circp=2\hat{m}B_{\bo}\,,\no[1mm]
\circkaon&=&2e^{2}\cf+(\mh+m_{s})B_{\bo}\,,\no
\circkanull\hspace{1mm}&=&\circk=(\mh+m_{s})B_{\bo}\,,\no[1mm]
\circsqm_{\eta}\hspace{3mm}&=&\frac{2}{3}(\mh+2m_{s})B_{\bo}\,.
\eqn
This decomposition depends on the convention adopted for the quark
self-energies. The effect of this ambiguity is small and neglected in this
article.
The correction to Dashen's theorem at order $e^2m_{q}$ is denoted by
\beq\label{nota}
\deltaka-\deltapi=
(\mkaon-\mkanull)-(\mpi-\mpin)\hspace{1.5cm} \mbox{
for}\hspace{5mm}m_{u}=m_{d}=\mh.
\eq
We calculate the Fourier transform of the two-point function of the mesons to
one loop,
keeping  the fields associated with the Gell-Mann matrices $\lambda_{a}$
instead of the physical ones. For the charged fields it has a cut at $p^2
=M_{a}^2$,
\beqn\label{twoch}
\lefteqn{i\int d^{4}x\;e^{ipx}<\bo|T\phix_{a}(x)\phix_{a}(\bo)e^{i\int d^{4}y
\left\{\laqi(y)+\laqc(y)\right\}}|\bo>|_{to\;one\;loop}}\no[2mm]
&=&\frac{\tilde{Z}_{a}(M_{a}^2)}{M^{2}_{a}
(1-p^{2}/M^{2}_{a})^{1+e^2 f_{a}(p^2)}}
+\cdots\hspace{1.5cm}a=1,2,4,5
\eqn
with
\beq
f_{a}(p^2)=\frac{1}{8\pi^{2}}\left(1+\frac{M_{a}^2}{p^{2}}\right),
\eq
whereas for the neutral fields it contains a pole,
\beqn
\lefteqn{i\int d^{4}x\;e^{ipx}<\bo|T\phix_{a}(x)\phix_{a}(\bo)e^{i\int d^{4}y
\left\{\laqi(y)+\laqc(y)\right\}}|\bo>|_{to\;one\;loop}}\no[2mm]
&=&\frac{Z_{a}(M_{a}^2)}{M_{a}^2-p^{2}}
+\cdots\hspace{1.5cm}a=3,6,7,8.
\eqn
$\laqi$ represents the interaction part of the lagrangian $\laq$,
$Z_{a}(M_{a}^2)$ and $\tilde{Z}_{a}(M_{a}^2)$ are due to the field
renormalization
and the ellipses denote terms which are not relevant for the mass-shifts at
order $e^{2}m_{q}$.
The contributions to the two-point function are shown in
fig. 1 : (b) and (c) are one-photon loops. The first contributes to the
two-point function of the charged fields only, the latter vanishes in
dimensional regularization. (d) is the tadpole with a four-meson coupling
and (e) represents the contributions from the lagrangian $\laqc$.
The results for the different mesons
are explicitly given in Appendix B. Here we consider only the difference
$\deltaka-\deltapi$,
\beqn\label{corrda}
\deltaka-\deltapi&=&-e^{2}\frac{1}{16\pi^{2}}\left[3\ck\ln\frac{\ck}{\mu^{2}}
-3\cp\ln\frac{\cp}{\mu^{2}}-4(\ck-\cp)\right]\no[1mm]
&&-e^{2}\frac{1}{8\pi^{2}}\cfy\left[\ck\ln\frac{\ck}{\mu^{2}}
-\cp(2\ln\frac{\cp}{\mu^{2}}+1)\right]\no[1mm]
&&-16e^{2}\cfy L_{5}^{r}(\mu)(\ck-\cp)+R_{\pi}(\mu)\cp+R_{K}(\mu)\ck\no[3mm]
&&+O(e^{2}m_{q}^2),
\eqn
where $L_{5}^{r}(\mu)$ and $R_{\pi,K}$ are the contributions from $\laqc$,
\beqn
R_{\pi}&=&\frac{2}{3}e^{2}\left(6K^{r}_{3}-3K^{r}_{4}
+2K^{r}_{9}-10K^{r}_{10}-12K^{r}_{11}\right),\no
R_{K}&=&-\frac{4}{3}e^{2}\left(K^{r}_{5}+K^{r}_{6}
-6K^{r}_{10}-6K^{r}_{11}\right).
\eqn
At the end we need $L^{r}_{5}$ and seven coupling constants $K^{r}_{i}$ from
$\laqc$ to determine the correction to Dashen's theorem at order $e^2 m_q$.

\subsection{Numerical Results}

We put ${\rm F}_{\bo}$ equal to the physical pion decay constant,
${\rm F}_{\pi}=93.3\, \mbox{MeV}$ and
the masses of the mesons to $\hat{M}_{\pi}=135\,\mbox{MeV},
\;\hat{M}_{K}=495\, \mbox{MeV}$. C can be expressed as an integral over the
difference of
the vector and axial vector spectral functions as established by Das,
Guralnik, Mathur, Low and Young \cite{das}. If we consider the low-lying
vector and axial-vector mesons $\rho$ and $A_1$, respectively, and use the
Weinberg sum rules \cite{sum} to eliminate the parameters of the $A_1$, we
arrive at \cite{ga82}
\beq
{\rm C}=\frac{3}{32\pi^{2}}M^{2}_{\rho}
{\rm F}_{\rho}^{2}\ln\left(\frac{{\rm F}_{\rho}^{2}}{{\rm F}_{\rho}^{2}
-{\rm F}_{\pi}^{2}}\right),
\eq
where $M_{\rho}$ is the mass of the $\rho$ meson,
$M_{\rho}=770\,\mbox{MeV}$, and ${\rm F}_{\rho}$ denotes the $\rho$ decay
constant, ${\rm F}_{\rho}= 154\,\mbox{MeV}$ \cite{ecker}. Therefore
\beq
{\rm C}=61.1\times 10^{-6}\,(\mbox{GeV})^4.
\eq
At tree level, the difference of the squared pion masses is
\beqn
\deltacircpi&=&(\hat{M}_{\pi^{\pm}}+\hat{M}_{\pi^{\bo}})(\hat{M}_{\pi^{\pm}}
-\hat{M}_{\pi^{\bo}})\no
&=&2M_{\pi}\times 4.8\,\mbox{MeV},
\eqn
which is very close to the experimental value,
$(M_{\pi^{\pm}}-M_{\pi^{0}})_{exp.}=4.6\,\mbox{MeV}$ \cite{data}.
The spectral functions can also
be extracted from data of the $\tau$-decay \cite{peccei}. However, this
method contains uncertainties because the spectral functions can
only be evaluated for momenta $p^{2}\leq 2(\mbox{GeV})^{2}$ and therefore we
keep
our approach above. $L^{r}_{5}(\mu)$ is taken from table 1 of ref.\cite{ecker}.
Its values at the scale points
$\mu=(0.5\,\mbox{GeV};\;M_{\rho};\;1\,\mbox{GeV})$ are
\beq
L_{5}^{r}(\mu)=(2.4;\,1.4;\,0.8)\pm 0.5\times 10^{-3}.
\eq
These
values lead to the following result for the squared mass differences, using
again the three
scales $\mu=(0.5\,\mbox{GeV};\;M_{\rho};\;1\,\mbox{GeV})$,
\beq\label{diff}
\deltaka-\deltapi=(-0.26;\,0.52;\,0.99)\pm 0.13\times 10^{-3}
(\mbox{GeV})^{2}\;
+R_{\pi}\cp+R_{K}\ck.
\eq
There is a strong dependence on the scale $\mu$ for all the terms. Of course,
this scale-dependence cancels in the full expression.\newline
Next we calculate the ratio $(\deltaka/\deltapi)$.
For this purpose we put $\deltapi=(\mpi-\mpin)_{exp.}$ and arrive at
\beq\label{ratio}
\frac{\deltaka}{\deltapi}=(0.79;\,1.42;\,1.80)\pm 0.11
+\frac{R_{\pi}\cp+R_{K}\ck}{\deltapi}.
\eq
In $\deltapi$ we can estimate the coupling constant $K_{8}^r(\mu)$ if we
neglect the unknown contributions from $\laqc$ proportional to $\cp$ (see
Appendix B),
\beqn\label{k10}
\deltapi&=&2e^{2}\cf -e^{2}\frac{1}{16\pi^{2}}\cp\left(3\ln\frac{\cp}
{\mu^{2}}-4\right)\no[1mm]
&&-e^{2}\frac{1}{8\pi^{2}}\cfy\left[\cp\left(3\ln\frac{\cp}
{\mu^{2}}+1\right)+\ck\ln\frac{\ck}{\mu^{2}}\right]\no[2mm]
&&-16e^{2}\cfy\left[(\cp+2\ck)L_{4}^{r}+\cp L_{5}^{r}
\right]+8e^{2}\ck K_{8}^{r}+O(e^2\cp).
\eqn
The values of $L_{4}^{r}(\mu)$ are at the three scalepoints
$\mu=(0.5\,\mbox{GeV};\;M_{\rho};\;1\,\mbox{GeV})$,
\beq
L_{4}^{r}(\mu)=(0.1;\;-0.3;\;-0.5)\pm 0.5\times 10^{-3}\,.
\eq
We put $\deltapi$ equal to the observed value and $K_{8}^r(\mu)$ becomes
\beq
K_{8}^r(\mu)=-(1.0;\;4.0;\;5.6)\pm1.7\times 10^{-3}\;.
\eq
$K_{8}^r$ is of the same order as the coefficients $L_{i}^r$ \cite{ga85} and as
it was expected by the so-called
naive chiral power counting \cite{georgi}. Note that $K_{8}^r(\mu)$ has not
the usual scale-dependence implied by (\ref{muk}), because we have dropped
scale-dependent terms in (\ref{k10}). Now we assume that the $K_{i}^r$'s
which contribute to $R_{\pi}$ and $R_{K}$ have the upper limit
\beq\label{ki}
|K_{i}|\stackrel{<}{\sim}\frac{1}{16\pi^{2}}=6.3\times10^{-3}\,.
\eq
The contributions from $R_{\pi}\cp+R_{K}\ck$ to the squared mass difference
$\deltaka-\deltapi$ is smaller than
\beq
|R_{\pi}\cp+R_{K}\ck|\;\stackrel{<}{\sim}\;2.6\times 10^{-3}
(\mbox{GeV})^{2}\;,
\eq
which is a large number. Similarly the contribution to the mass ratio
$(\deltaka/\deltapi)$ is
\beq
\frac{|R_{\pi}\cp+R_{K}\ck|}{\deltapi}\;\stackrel{<}{\sim}\;2.1\;.
\eq
On the basis of (\ref{ki}), large corrections to Dashen's theorem can
therefore not be excluded.

\subsection{Independent Approach}

We give a further estimate by looking at the QCD mass difference of the kaon.
In this calculation we put $m_{d}-m_{u}\neq 0$, but neglect corrections of
order $e^2(m_{d}-m_{u})$.
The observed squared mass difference of the kaon can be divided into two parts,
\beq\label{expka}
(\mkaon-\mkanull)_{exp.}=(\mkaon-\mkanull)_{QCD}+(\mkaon-\mkanull)_{e.m.}\,.
\eq
The same is true for the pions, but the QCD term is of $O[(m_{d}-m_{u})^2]$,
which is very small, numerically $(M_{\pi^{\pm}}-M_{\pi^{\bo}})_{QCD}=
0.17\,\mbox{MeV}$ \cite{ga85}. We put therefore
\beq\label{exppi}
(\mpi-\mpin)_{exp.}=(\mpi-\mpin)_{e.m.}.
\eq
The electromagnetic part of the squared mass difference using the notation of
(\ref{nota}) becomes
\beq
\deltaka-\deltapi=(\deltaka-\deltapi)_{exp.}-(\deltaka)_{QCD}
+O(e^{2}m_{q}^2).
\eq
The experimental part is $(\deltaka-\deltapi)_{exp.}
=-5.25\times10^{-3}(\mbox{GeV})^{2}$ \cite{data}.
We refer to Leutwyler \cite{leuti} for the calculation of
$(\deltaka)_{QCD}$ without using Dashen's theorem,
\beq
(\deltaka)_{QCD}=-(\ck-\cp)\frac{m_{d}-m_{u}}{m_{s}-\mh}
\left[1+\Delta_{M}+O(m_{q}^{2})\right].
\eq
$R=(m_{s}-\mh)/(m_{d}-m_{u})$ is the ratio determined from the mass
splittings in the baryon octet and from $\rho-\omega$ mixing \cite{ga82} with
the value $R=43.7\pm 2.7$. One way to obtain the correction $\Delta_{M}$
is an estimate of $\eta-\eta^{\prime}$ mixing. The phenomenological
informations \cite{ga85,dono} indicate that the mixing angle is somewhere
between
20 and 25 degrees. The range $|\theta_{\eta\eta^{\prime}}|=22^{\circ}\pm
4^{\circ}$ corresponds to a value of $\Delta_{M}=0.0\pm 0.12$. Inserting these
uncertainties into the equation above, the QCD squared mass difference of
the kaon becomes
\beq
(\deltaka)_{QCD}=-(5.19\pm 1.01)\times 10^{-3} (\mbox{GeV})^{2},
\eq
and the electromagnetic part of the squared mass difference is therefore (we
neglect the errors in the experimental values)
\beq\label{small}
\deltaka-\deltapi=-(0.06\pm 1.01)\times 10^{-3} (\mbox{GeV})^{2}
\eq
with a large uncertainty for a small value. Similarly the mass ratio
$(\deltaka/\deltapi)$ is
\beq
\frac{\deltaka}{\deltapi}=0.95\pm0.81\,.
\eq

\subsection{Comparison with Other Results}

Maltman and Kotchan \cite{maltmann} used the effective lagrangian $\laq$ and
calculated the one-loop
contributions to the squared mass differences $\Delta M_{\pi,K}^{2}$. They
consider terms proportional to $e^{2}\cp\ln
(\cp/\mu^2)$ and $e^{2}\ck\ln(\ck/\mu^2)$ at the scale point
$\mu=1.1\,\mbox{GeV}$. The scale independent terms $e^{2}\cp$ and $e^{2}\ck$
as well as the contributions from the effective lagrangian of order $p^4$ are
neglected. Their result is
\beq
(\deltaka-\deltapi)_{log}=0.76\times10^{-3}(\mbox{GeV})^{2},
\eq
which agrees with the contribution from the logarithms in our calculation
at $\mu=1.1\,\mbox{GeV}$,
\beq
(\deltaka-\deltapi)_{log}=0.77\times10^{-3}(\mbox{GeV})^{2}.
\eq
The authors expect that the ratio
$(\deltaka/\deltapi)$ is in the range (at this scale point)
\beq
\frac{\deltaka}{\deltapi}=1.44\pm 0.20.
\eq
Using again $\deltapi=(\deltapi)_{exp.}$, this corresponds to
\beq
\deltaka-\deltapi=(0.55\pm 0.25)\times 10^{-3}({\rm GeV})^2 .
\eq
Recently two other estimates have been made by Donoghue, Holstein and Wyler
\cite{wyler} and Bijnens \cite{bijnens}. The first work is based on chiral
symmetry and vector meson dominance. Their result shows a
total ratio which is rather large,
\beq\label{large}
\frac{\deltaka}{\deltapi}=1.8\;.
\eq
The value for $\deltapi$ they obtain at this order is about 20\% higher than
the
experimental value,
\beqn\label{pimass}
\mpi-\mpin=2M_{\pi}\times 5.6\,\mbox{MeV}.
\eqn
Inserting (\ref{pimass}) into the
ratio above (\ref{large}), the electromagnetic part of the squared mass
difference becomes
\beq
\deltaka-\deltapi=1.23\times10^{-3}(\mbox{GeV})^{2}.
\eq
This result is in good correspondence with  the one obtained by Bijnens
\cite{bijnens},
\beq
\deltaka-\deltapi=(1.3\pm 0.4)\times10^{-3}(\mbox{GeV})^{2},
\eq
who calculated this value by splitting the contributions to the masses into two
parts. A long distance part is estimated using the $(1/N_{c})$-approach
and a short distance contribution is determined in terms of the
coupling constants $L_{i}^r$ of the lagrangian $\laqc$.\newline
The results we listed in this section show rather large corrections to Dashen's
theorem. They coincide only partly with the result we obtained in
(\ref{small}), but they could be in agreement with the estimate on the
basis of (\ref{ki}). In order to establish them, we would have to
numerically evaluate the relevant coefficients $K_{i}^{r}$ of $\laqc$.

\section{Summary}

{\bf 1.}\hspace{4mm}We have evaluated the divergent part of the
generating functional of CHPT including virtual photons to one loop. We have
calculated the structure
of the local action at order $p^4$ with 29 coupling constants
$(L_1,\ldots,L_{10};H_1,\,H_2;K_1,\ldots,K_{17})$, which have in general a
single pole in $d=4$ dimensions.\newline
\newline
{\bf 2.}\hspace{4mm}In a next step we have calculated the one-loop
contribution to the meson masses in the isospin limit $m_{u}=m_{d}=\hat{m}$
and extracted the
correction to Dashen's theorem \cite{dashen},
\beqn
\deltaka-\deltapi&=&-e^{2}\frac{1}{16\pi^{2}}\left[3\ck\ln\frac{\ck}{\mu^{2}}
-3\cp\ln\frac{\cp}{\mu^{2}}-4(\ck-\cp)\right]\no[1mm]
&&-e^{2}\frac{1}{8\pi^{2}}\cfy\left[\ck\ln\frac{\ck}{\mu^{2}}
-\cp(2\ln\frac{\cp}{\mu^{2}}+1)\right]\no[1mm]
&&-16e^{2}\cfy L_{5}^{r}(\mu)(\ck-\cp)+R_{\pi}(\mu)\cp+R_{K}(\mu)\ck\no[3mm]
&&+O(e^{2}m_{q}^2),
\eqn
where $\Delta M_{P}^{2}=M_{P^{\pm}}^{2}-M_{P^{\bo}}^{2}$ with $P$ the
pseudoscalar meson in question.
$L_{5}^{r}(\mu)$ and $R_{\pi,K}$ are the contributions from $\laqc$,
\beqn
R_{\pi}&=&\frac{2}{3}e^{2}\left(6K^{r}_{3}-3K^{r}_{4}
+2K^{r}_{9}-10K^{r}_{10}-12K^{r}_{11}\right),\no
R_{K}&=&-\frac{4}{3}e^{2}\left(K^{r}_{5}+K^{r}_{6}
-6K^{r}_{10}-6K^{r}_{11}\right).
\eqn
At the three
scale points $\mu=(0.5;\;0.77;\;1)\,\mbox{GeV}$ we got numerically
\beqn\label{summi}
\deltaka-\deltapi=(-0.26;\,0.49;\,0.99)\pm 0.13\times 10^{-3}
(\mbox{GeV})^{2}\;
+R_{\pi}\cp+R_{K}\ck.
\eqn
We assumed the upper limit of $R_{\pi}\cp+R_{K}\ck$ to be
\beq
|R_{\pi}\cp+R_{K}\ck|\;\stackrel{<}{\sim}\;2.6\times 10^{-3}(\mbox{GeV})^{2}\;,
\eq
from which we can not exclude large corrections to Dashen's theorem.
Note that in the numerical part in (\ref{summi}) the correction due to
$L_{5}^r(\mu)$
is already included.\newline
\newline
{\bf 3.}\hspace{4mm}In addition, we have
given an estimate of $\deltaka-\deltapi$ on the basis of ref.\cite{leuti},
\beq\label{summii}
\deltaka-\deltapi=-(0.06\pm 1.01)\times 10^{-3} (\mbox{GeV})^{2},
\eq
a small number with a large uncertainty.\newline
\newline
{\bf 4.}\hspace{4mm}We have compared our values to the
results in the literature \cite{maltmann,wyler,bijnens}. We found that they
coincide only partly with the result (\ref{summii}), but they could be well in
agreement
with the estimate in (\ref{summi}). In order to complete the expression for
$\deltaka-\deltapi$
we need the numerical evaluation of $R_{\pi,K}$, i.e  the calculation of the
finite parts of the coupling constants
$K_{i}$. This is beyond the scope of this work.

\subsection*{Acknowledgements}

I am grateful to J\"{u}rg
Gasser for many discussions and for reading this
manuscript carefully. Also I thank  H. Neufeld who pointed out, that my
original lagrangian $\laqc$ contains an overcomplete set of counterterms, J.
Bijnens for correspondence concerning
effective lagrangians, Urs B\"urgi, who helped to draw the
Feynman graphs, and Robert Baur for pointing out a numerical error in the
coefficients $\Sigma_{i}\,$, that he had worked out
independently.

\appendix
\section {Matrix Relations}

For the derivation of the effective lagrangian $\laqc$ at order $p^4$, we used
the equation of motion obeyed by $\bar{U}$,
\beqn
\lefteqn{d^{\mu}d_{\mu}U^{+} U - U^{+} d^{\mu}d_{\mu}U + (U^{+}\chi -
\chi^{+}U)}\no
&& + 4\cf (U^{+}QUQ - QU^{+}QU) -\frac{1}{3}<U^{+}\chi - \chi^{+}U>=0.
\eqn
The Cayley - Hamilton theorem for a $3\times 3$-matrix $A$,
\beqn
\lefteqn{A^3-<A>A^2+\frac{1}{2}\left\{<A>^2-<A^2>\right\}A}\no
&&-\frac{1}{6}\left\{<A>^3-3<A><A^2>+2<A^3>\right\}{\rm {\bf 1}}=0,
\eqn
implies a trace identity for $3\times 3$-matrices $A,B,C$ and $D$,
\beqn
<ABCD>&=&-<ABDC>-<ACBD>-<ACDB>\no
&&-<ADBC>-<ADCB>+<AB><CD>\no
&&+<AC><BD>+<AD><BC>+<A><BCD>\no
&&+<A><BDC>+<B><ACD>+<B><ADC>\no
&&+<C><ABD>+<C><ADB>+<D><ABC>\no
&&+<D><ACB>-<A><B><CD>\no
&&-<A><C><BD>-<A><D><BC>\no
&&-<B><C><AD>-<B><D><AC>\no
&&-<C><D><AB>+<A><B><C><D>.
\eqn
Furthermore we used the observation that $Q^2$ can be written as
\beq
Q^2=\frac{e}{3}Q + \frac{2e^2}{9}{\bf 1}.
\eq
We confirmed the 17 independent terms $K_i<\cdots>$ $(i=1,2,\ldots,17)$ in
$\laqc$ calculating the contributions from tree graphs at order $p^4$ to
\begin{itemize}
\item the constant term $e^4{\bf 1}$,
\item the masses $\mpi,\mpin$ and $M^{2}_{\eta}$,
\item the scattering amplitudes at order $e^2 m_q$
\beqn
\pi^{+}\pi^{-}&\rightarrow&\pi^{+}\pi^{-},\no
\pi^{+}\pi^{-}&\rightarrow&K^{\bo}\bar{K}^{\bo},\no
K^{+}K^{-}&\rightarrow&K^{\bo}\bar{K}^{\bo},
\eqn
\item the scattering amplitudes at order $e^4$
\beqn
\pi^{+}\pi^{-}&\rightarrow&\pi^{+}\pi^{-},\no
\pi^{+}\pi^{-}&\rightarrow&K^{\bo}\bar{K}^{\bo},
\eqn
\item the matrix elements
\beqn
\lefteqn{<\pi^{+}|\bar{u}\gamma_5 d|\pi^{+}\pi^{-}>,\hspace{5mm}
<\pi^{+}|\bar{u}\gamma_5 d|\eta\eta>,}\no
\lefteqn{<0|\bar{u}\gamma_{\mu}\gamma_5 d|\pi^{-}>,\hspace{5mm}
<0|T\,\bar{u}(x)\gamma^{\mu}\gamma_5 d(x)\bar{d}(y)\gamma_{\mu}\gamma_5
u(y)|0>,}\no
\lefteqn{<\pi^{+}|T\,\bar{u}(x)\gamma^{\mu}d(x)\bar{d}(y)
\gamma_{\mu}u(y)|\pi^{+}>,\hspace{5mm}
<0|T\,\bar{u}(x)\gamma^{\mu}d(x)\bar{d}(y)\gamma_{\mu}u(y)|0>.}
\eqn
\end{itemize}

\section{The Formal Expressions for the Masses at Order $e^{2}m_{q}$}

The photon loop (see fig.1) gives a contribution to the masses of
the charged particles,
\beq
\delta M_{C^{\pm}}^{2}|_{\gamma-loop}= -e^{2}\frac{1}{16\pi^{2}}M^{2}_{C}
\left(3\ln\frac{M^{2}_{C}}{\mu^{2}}-4\right),
\eq
where $C$ stands for $\pi$ and $K$, respectively.\newline
{}From the tadpole, the masses of the charged particles get a change of the
form
\beq
\delta M_{C^{\pm}}^{2}|_{tadpole}=-e^{2}\frac{1}{8\pi^{2}}\cfy
\left[A_{C^{\pm}}\cp\ln\frac{\cp}{\mu^{2}}+B_{C^{\pm}}\ck\ln
\frac{\ck}{\mu^{2}}\right] ,
\eq
with $(A_{\pi^{\pm}},\,B_{\pi^{\pm}})=(2,\,1)$ and
$(A_{K^{\pm}},\,B_{K^{\pm}})=(1,\,2)$. The contributions to the masses of
$\pi^{\bo}$ and $\eta$ are
\beq
\delta M_{P}^{2}|_{tadpole}=e^{2}\frac{1}{48\pi^{2}}
\cfy\left[\alpha_{P}\cp\left(\ln\frac{\cp}{\mu^{2}} +1\right)
+\beta_{P}\ck\left(
\gamma_{P}\ln\frac{\ck}{\mu^{2}}+1\right)\right],
\eq
with the coefficients
$(\alpha_{\pi^{\bo}},\,\beta_{\pi^{\bo}},\,\gamma_{\pi^{\bo}}) =(6,\,0,\,0)$
and $(\alpha_{\eta},\,\beta_{\eta},\,\gamma_{\eta})
=(-2,\,1,\,4)$. The masses of the neutral kaons do not get contributions
from the loops. At order $e^{2}m_{q}$, the mass-shift due to $\laqc$  involves
in general only terms with the couplings $K_i$, but for
the charged particles, where $L_{4}$ and $L_{5}$ enter as well,
\beqn
\delta M_{C^{\pm}}^{2}|_{\laqc}=-16e^{2}\cfy
\left[(\cp+2\ck)L_{4}^{r}(\mu) + M_{C}^{2}L_{5}^{r}(\mu)\right]+\cdots.
\eqn
The term proportional to $L_{4}^{r}(\mu)$ drops out in the difference
$\deltaka-\deltapi$. Explicitly, the masses are at $O(e^{2}m_{q})$,
\newpage
\baselineskip 8mm
\beqn
\mpi&=&\circpi -e^{2}\frac{1}{16\pi^{2}}\cp\left(3\ln\frac{\cp}
{\mu^{2}}-4\right)\no
&&\hspace{8mm}-e^{2}\frac{1}{8\pi^{2}}\cfy\left[2\cp\ln
\frac{\cp}{\mu^{2}}+\ck\ln\frac{\ck}{\mu^{2}}\right]\no
&&\hspace{8mm}-16e^{2}\cfy\left[(\cp+2\ck)L_{4}^{r}+\cp L_{5}^{r}
\right]\no
&&\hspace{8mm}-\frac{4}{9}e^{2}\cp\left(6K_{1}^{r}+6K_{2}^{r}
+5K_{5}^{r}+5K_{6}^{r}-6K_{7}^{r}\right.\no
&&\hspace{2.3cm}\left.-15K_{8}^{r}
-5K_{9}^{r}-23K_{10}^{r}-18K_{11}^{r}\right)\no
&&\hspace{8mm}+8e^{2}\ck K_{8}^{r}\;,\\
\no
\mpin&=&\circpin +e^{2}\frac{1}{8\pi^{2}}\cfy\cp\left(\ln\frac{\cp}
{\mu^{2}}+1\right)\no
&&\hspace{7mm}-\frac{2}{9}e^{2}\cp\left(12K_{1}^{r}+12K_{2}^{r}-18K_{3}^{r}
+9K_{4}^{r}+10K_{5}^{r}\right.\no
&&\hspace{2.3cm}\left.+10K_{6}^{r}-12K_{7}^{r}-12K_{8}^{r}
-10K_{9}^{r}-10K_{10}^{r}\right),\\
\no
\mkaon&=&\circkaon -e^{2}\frac{1}{16\pi^{2}}\ck\left(3\ln\frac{\ck}
{\mu^{2}}-4\right)\no
&&\hspace{9mm}-e^{2}\frac{1}{8\pi^{2}}\cfy\left[\cp\ln
\frac{\cp}{\mu^{2}}+2\ck\ln\frac{\ck}{\mu^{2}}\right]\no
&&\hspace{1cm}-16e^{2}\cfy\left[(\cp+2\ck)L_{4}^{r}+\ck L_{5}^{r}
\right]\no
&&\hspace{9mm}+\frac{4}{3}e^{2}\cp\left(3K_{8}^{r}
+K_{9}^{r}+K_{10}^{r}\right)\no
&&\hspace{9mm}-\frac{4}{9}e^{2}\ck\left(6K_{1}^{r}+6K_{2}^{r}
+5K_{5}^{r}+5K_{6}^{r}-6K_{7}^{r}\right.\no
&&\hspace{2.5cm}\left.-24K_{8}^{r}
-2K_{9}^{r}-20K_{10}^{r}-18K_{11}^{r}\right),\\
\no
\mkanull&=&M_{\bar{K}^{\bo}}^{2}\no
&=&\circkanull -\frac{8}{9}e^{2}\ck\left(3K_{1}^{r}+3K_{2}^{r}
+K_{5}^{r}+K_{6}^{r}\right.\no
&&\hspace{2.5cm}\left.-3K_{7}^{r}-3K_{8}^{r}
-K_{9}^{r}-K_{10}^{r}\right),\\
\no
M_{\eta}^{2}&=&\circsqm_{\eta} -e^{2}\frac{1}{48\pi^{2}}\cfy
\left[2\cp\left(\ln\frac{\cp}{\mu^{2}}+1\right)
-\ck\left(4\ln\frac{\ck}{\mu^{2}}+1\right)\right]\no
&&\hspace{6mm}+\frac{4}{9}e^{2}\cp\left(K_{9}^{r}+K_{10}^{r}\right)\no
&&\hspace{6mm}-\frac{2}{9}e^{2}M^{2}_{\eta}\left(12K_{1}^{r}+12K_{2}^{r}
-6K_{3}^{r}+3K_{4}^{r}+6K_{5}^{r}\right.\no
&&\hspace{2.1cm}\left.+6K_{6}^{r}-12K_{7}^{r}-12K_{8}^{r}
-4K_{9}^{r}-4K_{10}^{r}\right).
\eqn
The squared mass differences of the pions and the kaons are
\beqn
\deltapi&=&\mpi-\mpin\no
&=&2e^{2}\cf -e^{2}\frac{1}{16\pi^{2}}\cp\left(3\ln\frac{\cp}
{\mu^{2}}-4\right)\no
&&\hspace{1cm}-e^{2}\frac{1}{8\pi^{2}}\cfy\left[\cp\left(3\ln\frac{\cp}
{\mu^{2}}+1\right)+\ck\ln\frac{\ck}{\mu^{2}}\right]\no
&&\hspace{1cm}-16e^{2}\cfy\left[(\cp+2\ck)L_{4}^{r}+\cp L_{5}^{r}
\right]\no
&&\hspace{1cm}+2e^{2}\cp\left(-2K_{3}^{r}+K_{4}^{r}
+2K_{8}^{r}+4K_{10}^{r}+4K_{11}^{r}\right)\no
&&\hspace{1cm}+8e^{2}\ck K_{8}^{r}\;,
\eqn
\beqn
\deltaka&=&\mkaon-\mkanull\no
&=&2e^{2}\cf -e^{2}\frac{1}{16\pi^{2}}\ck\left(3\ln\frac{\ck}
{\mu^{2}}-4\right)\no
&&\hspace{1cm}-e^{2}\frac{1}{8\pi^{2}}\cfy\left[\cp\ln\frac{\cp}
{\mu^{2}}+2\ck\ln\frac{\ck}{\mu^{2}}\right]\no
&&\hspace{1cm}-16e^{2}\cfy\left[(\cp+2\ck)L_{4}^{r}+\ck L_{5}^{r}
\right]\no
&&\hspace{1cm}+\frac{4}{3}e^{2}\cp\left(3K_{8}^{r}
+K_{9}^{r}+K_{10}^{r}\right)\no
&&\hspace{1cm}-\frac{4}{3}e^{2}\ck\left(K_{5}^{r}
+K_{6}^{r}-6K_{8}^{r}-6K_{10}^{r}
-6K_{11}^{r}\right).
\eqn
And finally the difference $\deltaka-\deltapi$ is
\beqn
\deltaka-\deltapi&=&-e^{2}\frac{1}{16\pi^{2}}\left[3\ck\ln\frac{\ck}{\mu^{2}}
-3\cp\ln\frac{\cp}{\mu^{2}}-4(\ck-\cp)\right]\no
&&-e^{2}\frac{1}{8\pi^{2}}\cfy\left[\ck\ln\frac{\ck}{\mu^{2}}
-\cp(2\ln\frac{\cp}{\mu^{2}}+1)\right]\no
&&-16e^{2}\cfy L_{5}^{r}(\mu)(\ck-\cp)+R_{\pi}(\mu)\cp
+R_{K}(\mu)\ck\, ,
\eqn
where $R_{\pi,K}$ are,
\beqn
R_{\pi}&=&\frac{2}{3}e^{2}\left(6K^{r}_{3}-3K^{r}_{4}
+2K^{r}_{9}-10K^{r}_{10}-12K^{r}_{11}\right),\no
R_{K}&=&-\frac{4}{3}e^{2}\left(K^{r}_{5}+K^{r}_{6}
-6K^{r}_{10}-6K^{r}_{11}\right).
\eqn

\end{document}